\documentclass{article}
\usepackage{arxiv}

\usepackage[utf8]{inputenc} 
\usepackage[T1]{fontenc}    
\usepackage{hyperref}       
\usepackage{url}            
\usepackage{booktabs}       
\usepackage{amsfonts}       
\usepackage{nicefrac}       
\usepackage{microtype}      
\usepackage{lipsum}		
\usepackage{graphicx}
\usepackage{natbib}
\usepackage{doi}
\usepackage{amsmath}

\usepackage{amssymb,amsfonts,amsthm,bm,bbm}
\usepackage{algorithm}
\usepackage{soul}
\usepackage{algorithmic}
\usepackage{array}
\usepackage{graphicx}
\usepackage{booktabs}
\usepackage{multirow}
\usepackage{pifont}
\usepackage{tcolorbox}
\usepackage{enumitem}
\usepackage{multicol}
\usepackage{amsmath,amssymb,amsfonts,amsthm}
\usepackage{graphicx}
\usepackage{comment}
\usepackage{diagbox}
\usepackage{bigfoot} 
\usepackage{thmtools}
\usepackage{thm-restate}
\usepackage{soul}
\DeclareMathOperator*{\argmax}{argmax}

\usepackage{algorithm,algorithmic}
\usepackage{subcaption}
\usepackage{url}

\newtheorem{proposition}{Proposition}
\newtheorem{claim}{Claim}
\newtheorem{theorem}{Theorem}
\newtheorem{lemma}[theorem]{Lemma}

\usepackage{xcolor}
\newcommand{\proto}{\emph{QuickSync}}

\newcommand{\btc}{\emph{Bitcoin}}

\title{\proto: A Quickly Synchronizing PoS-Based Blockchain Protocol}

\date{February 13, 2023}	

\author{ 
    Shoeb Siddiqui \\
	Machine Learning Lab\\
	IIIT Hyderabad\\
	\texttt{shoeb.siddiqui@research.iiit.ac.in} \\
	\And
    \href{https://orcid.org/0000-0002-5662-0386}{\includegraphics[scale=0.06]{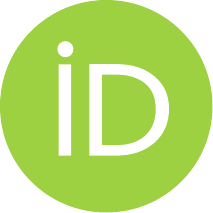}\hspace{1mm}Varul Srivastava} \\
	Machine Learning Lab\\
	IIIT Hyderabad\\
	\texttt{varul.srivastava@research.iiit.ac.in} \\
	\And
        Raj Maheshwari \\
	IIIT Hyderabad\\
	\texttt{raj.maheshwari@students.iiit.ac.in} \\
	\And
	\href{https://orcid.org/0000-0003-4634-7862}{\includegraphics[scale=0.06]{orcid.pdf}\hspace{1mm}Sujit Gujar} \\
	Machine Learning Lab\\
	IIIT Hyderabad\\
	\texttt{sujit.gujar@iiit.ac.in} \\
}

\renewcommand{\shorttitle}{QuickSync}

\hypersetup{
pdftitle={QuickSync},
pdfsubject={q-bio.NC, q-bio.QM},
pdfauthor={Shoeb Siddiqui, Varul Srivastava, Raj Maheshwari, Sujit Gujar},
pdfkeywords={Blockchains, Distributed Systems},
}

\begin{document}
\maketitle

\begin{abstract}
Blockchain technology has found applications for myriad use cases. Proof-of-Stake(PoS) based blockchain protocols have gained popularity due to their higher throughput and low carbon footprint when compared with Proof-of-Work blockchain protocols. The two major parts of blockchain protocols is selection of the next block proposer and selection of the longest chain. In the Proof-of-Stake (PoS) consensus mechanism, block publishers are selected based on their relative stake. However, PoS-based blockchain protocols may face vulnerability against Fully Adaptive Corruptions.

This paper proposes a novel PoS-based blockchain protocol, QuickSync, to achieve security against Fully Adaptive Corruptions while improving performance. Towards this, we propose a metric for each block: block power. This is a function of the block publisher's stake derived from a verifiable random function (VRF) output. We compute the chain power of a chain as the sum of block powers of all the blocks comprising the chain. The chain selection rule selects the chain with the highest chain power as the valid chain. Since block proposer is not selected upfront, this scheme is resilient to fully adaptive corruptions, which we also show formally. We also ensure that our block power mechanism is resistant to Sybil attacks. We prove the security of QuickSync by showing that it satisfies the common prefix, chain growth, and chain quality properties. We also show its resilience against different adversarial attack strategies. Our analysis demonstrates that QuickSync performs better than Bitcoin by order of magnitude on both transactions per second and time to finality. 

\end{abstract}

\section{Introduction}
\label{sec:intro}

A \emph{blockchain} is an append-only, secure, transparent, distributed ledger. It stores the data in \emph{blocks} connected through immutable cryptographic links, with each block extending exactly one previous block.
Introduced in Bitcoin \cite{nakamoto2008bitcoin}, blockchain is one of this century's most significant technological innovations. 
The underlying technical problem that Bitcoin solves through blockchain is \emph{byzantine fault tolerant} \emph{distributed consensus} in a decentralized system. 

In any blockchain system, 
delays in communication and adversarial attacks may cause \emph{forks} in the chain, creating ambiguity as to extend which block. The consensus protocol prevents these forks by selecting a node for publishing blocks using a \emph{block publisher selection mechanism} (BPSM). However, forks may still occur. To resolve these forks, we require a \emph{chain selection rule} (CSR) to choose one among them. The BPSM and CSR functionally characterize a blockchain protocol. Bitcoin blockchain protocol uses a \emph{Proof-of-Work} (PoW) based BPSM and \emph{longest chain} as the CSR. 
Typically, we measure a blockchain's performance with the two metrics: i) \emph{transactions per second} ($tps$), and ii) \emph{time for finality} ($t_f$), i.e., the time required to confirm a transaction.

Although a great innovation has certain challenges regarding efficiency and performance, Bitcoin consumes high electrical power to function. As of writing this paper, the Bitcoin network uses an estimated $266.3$ Ehash/s \cite{bitinfocharts}, and annual power consumption of $113$ TWh \cite{digiconomist}. Typically in Bitcoin $tps=4$ and $t_f=60$ minutes \cite{blockchain.com}. Due to the poor $tps$, it faces scalability issues, and the high $t_f$ deters usability and adoption.
Attempts such as \emph{GHOST} \cite{sompolinsky2013accelerating} aim to improve performance by using a different CSR. In GHOST, the BPSM is still PoW-based, and its power consumption remains high. Towards addressing high power consumption, researchers proposed \emph{Proof-of-Stake} (PoS) based BPSM. 

PoS-based blockchain protocols, such as Algorand \cite{gilad2017algorand}, and Casper \cite{buterin2017casper}, stochastically choose the \emph{selected block publisher} (SBP) with probability proportional to its \emph{relative stake}. This approach is also consistent with the expectation that nodes that are stakeholders would not want to endanger the system. The \emph{Ouroboros} protocols \cite{kiayias2017ouroboros,david2018ouroboros,badertscher2018ouroboros,kerber2018ouroboros, ouroboros_chronos} are popular PoS-based protocols, amongst others. Ouroboros is not only academically published\footnote{Some PoS protocols are not academically published, e.g. Casper} in a peer-reviewed forum, but is also the foundation for the crypto-currency \emph{Cardano} \cite{cardano}.

\emph{Ouroboros v1} (v1) \cite{kiayias2017ouroboros} achieves high $tps$ and better $t_f$ than Bitcoin. However, for a blockchain protocol to be completely secure, it must be immune to \emph{fully adaptive corruption}s (FACs), i.e., a \emph{dynamic adversary}. Ouroboros v1 is not immune to FACs.
\emph{Ouroboros Praos} (Praos) \cite{david2018ouroboros} uses a different BPSM. Praos does solve the security weakness but does not improve performance. While both \emph{Ouroboros Genesis} \cite{badertscher2018ouroboros} and \emph{Ouroboros Crypsinous} \cite{kerber2018ouroboros} protocols enhance the Ouroboros protocol, they too do not improve performance. Ouroboros Chronos \cite{ouroboros_chronos} fixes the clock synchronization problem and leverages PoS to establish a common notion of global time without a global clock.

In this work, we propose a novel blockchain protocol, \proto\, that is secure against FACs, and achieves slightly better $tps$, and improves on the $t_f$ by a factor of 3, as compared to Ouroboros v1. Essentially, it quickly synchronizes (resolves) the forks that arise. To build \proto, we employ the framework of the Ouroboros protocol. \proto\ differs from v1 and Praos in both the BPSM and the CSR.
The key idea is to propose a metric called \emph{block power} assigned to each block. Using this, we compute the \emph{chain power} of every competing chain as the sum of all block powers in that chain. We then establish the \emph{best chain} with the highest chain power from a given set of chains. All forks are thus trivially resolved, except for the ones generated by the adversary. Block power is a function of the node's privately computed \emph{Verifiable Random Function} (VRF) output dependent on the node's private key, seed randomness, and the slot counter. Since this VRF output is not revealed until the block is published, \proto\ is immune against Fully Adaptive Corruptions. Block power is also a function of the node's relative stake, enabling the PoS aspect of the protocol. We have designed our block power to be resistant to Sybil attacks.
As multiple nodes publish blocks simultaneously, it may seem that there will be several forks in \proto, as is the case in the other PoS protocols such as Praos. The key novelty here is that we resolve these forks immediately using block power.
Thus, \proto\ is in sharp contrast to other PoS-based mechanisms that \emph{do not differentiate} between the published blocks.

Researchers showed that any blockchain protocol satisfying the three properties: \emph{common-prefix}, \emph{chain-growth}, and \emph{chain-quality} implements a robust transaction ledger  \cite{garay2015bitcoin,kiayias2015speed,pass2017analysis}. We prove that \proto\ satisfies these three properties (Theorem \ref{theo:sr}) and ascertain that our protocol is immune to FACs (Proposition \ref{prop:fac}). We also examine our protocol for different attack strategies and show resistance to them.

\textbf{key Contributions.} In summary, the PoS-based blockchain protocol, \proto\ fixes the security weakness of Ouroboros v1 and performs better in terms of $tps$ and $t_f$ by about an order of magnitude as compared to Bitcoin.
\begin{itemize}
  \item We have developed a novel CSR mechanism through a Sybil-resistant function that we call block power. Our CSR mechanism is capable of an instant resolution of forks.
  \item We propose a simple and elegant PoS protocol that is secure against Fully Adaptive Corruptions yet highly efficient compared to other PoS-based protocols.
\end{itemize}

The rest of the paper is organized as follows. In Section \ref{sec:prelim}, we explain the relevant preliminaries. In Section \ref{ssec:op_oa} we discuss the intuition behind \proto. In Section \ref{sec:ourproto}, we describe our protocol, \proto. We discuss how to set the parameters for \proto\ in Section \ref{sec:disc}. In Section \ref{sec:rel}, we summarize the related work in this domain. Finally, we discuss future work and conclude the paper in Section \ref{sec:conc}. The detailed security analysis, analysis of attack strategies, and a comparative note are skipped due to space constraints. An anonymous version can be provided if required. 

\section{Preliminaries}
\label{sec:prelim}
In this section, we discuss the preliminary notions and key concepts required to build \proto. 
\subsection{General Concepts and Notation}
\label{ssec:notation}

\subsubsection{Blockchain}
A blockchain is a singular sequence of blocks connected by hash links.
A block is comprised of a \emph{block header} and \emph{block data}. The block header contains, amongst other things, the publisher's public key and the hash of the \emph{merkle tree root} of the block data, as well as the hash of the previous block. The block data contains the transactions or any other data that is to be added to the record. \emph{Block publisher} is the node that has built the block in consideration. We refer to the number of transactions per block as $tpb$.
A chain of blocks, $C$ referred to as chain, consists of an ordered set of blocks, where every block, $B^l; l>0$ is immutably linked to the block, $B^{l-1}$, $\forall B^l \in C$; where $l$ represents the ordinal number, and $B^0$ is the genesis block. The length of a chain, $C$, is denoted as $len(C)$, is the number of blocks in the chain excluding the genesis block. We say that the chain selected by the CSR has been \emph{adopted} and is \emph{held} by the node.

\subsubsection{Blockchain Protocol}

Each node is identified by a public key $pk$ and holds a master secret key $msk$. $\mathbb{S}$ is the set of all nodes, composed of $\mathbb{H}$, the set of all honest nodes (motivated by reward schemes) and, $\mathbb{A}$, the adversary. In a decentralized system, multiple nodes from $\mathbb{S}$ would be interested in writing the next block to the blockchain. We need to select a publisher for proposing each new block. This publisher is called the \emph{selected block publisher} (SBP). The SBP is selected through a \emph{block publisher selection mechanism}. Due to network delays and adversarial attacks, it is possible to have multiple versions of the chains, i.e., forks from the previously agreed state of the chain. While designing a blockchain protocol, there should be an implicit or explicit rule to resolve these forks. We refer to this as \emph{chain selection rule} (CSR). In summary, to design a blockchain protocol, we need to specify BPSM and CSR. 


\subsubsection{Proof-of-Stake Blockchain Protocols}
In \emph{Proof-of-Stake} (PoS) protocols, each node $n \in \mathbb{S}$ has influence proportional to the amount of \emph{relative stake}, the fraction of the total stake held by the node. This influence is expressed as the probability of being selected as a SBP. 
$r_h$ and $r_a$ denotes the relative stake of the honest nodes and the adversary respectively ($r_h,r_a \geq 0$, $r_h + r_a = 1$).
The relative stake that is active in the execution of the protocol at any given moment is $r^{active}\geq 0$.

\subsubsection{Forks}
A \emph{fork} in a blockchain is the case when two different blocks extend the same block. 
The adversary may attempt to create a fork, privately or publicly, with the intention to compromise the protocol. These forks enable an adversary to double spend. An essential part of blockchain protocols is to ensure that these forks are resolved quickly,
enabling relative finality with time.

\subsubsection{Relative Finality} To establish \emph{relative finality}, we define \emph{k-finality} (referred to as finality) as the property of a blockchain protocol. We say the protocol has finality with parameter $k$, if all the honest nodes can confirm a block $B$, once $k$ valid blocks have extended the chain after block $B$. Similar to Bitcoin and Ouroboros, we establish relative finality, i.e., the confirmation of a block by an honest node implies that with probability $1-\eta$ ($0<\eta<1$), no other honest node will ever in the future disagree with the confirmed block's placement in the ledger.

To violate k-finality, the adversary must show a chain, $C'$, that makes the honest nodes replace their CSR selected chain, $C$, where $C$ and $C'$ differ by atleast $k$ blocks.

\subsubsection{Performance metrics} We define \emph{transactions per second} ($tps$) of a protocol, as the maximum number of transactions that the protocol can add to its record every second.
We define \emph{time to finality} ($t_f$) as the time it takes for the protocol to confirm a block once published, given a certain assurance level of $1-\eta$, $0<\eta<1$. Please note that $t_f=k\cdot t_{sl}$, where $k$ is the \emph{common prefix parameter} and $t_{sl}$ is the slot length\footnote{fixed units of predefined time length. Defined in Section~\ref{ssec:time-and-slots}} in units of time.

\subsubsection{Requisites for a Blockchain Consensus Mechanism}
\label{ssec:reqbp}
A blockchain is, in essence, a transaction ledger. For a protocol to implement a robust transaction ledger, it must satisfy two properties \cite{garay2015bitcoin}:
\begin{itemize}
    \item \emph{liveness}: Once a node broadcasts a transaction, it will eventually be confirmed by getting included in the transaction ledger.
    \item \emph{persistence}: Once an honest node confirms a transaction, then all the other honest nodes agree with its placement in the ledger.
\end{itemize}

The authors of \cite{kiayias2015speed,pass2017analysis} showed that liveness and persistence are equivalent to \emph{common prefix}, \emph{chain growth} and, \emph{chain quality} for any blockchain protocol.
For a PoS-based blockchain protocol that utilizes slots, these three properties are defined as follows in \cite{kiayias2017ouroboros}. 
\begin{itemize}
    \item \emph{Common prefix}: We say that the protocol satisfies common prefix property with parameter $k$, if given the adopted chains $C_1$ and $C_2$ of any two honest nodes $n_1$ and $n_2$ at slots $l_1$ and $l_2$ respectively such that $l_1<l_2$, then by removing $k$ blocks from the end of $C_1$ we should get the prefix of $C_2$.

    \item \emph{Chain growth}: We say that the protocol satisfies chain growth with parameter $\zeta$, if given the adopted chains $C_1$ and $C_2$ of any honest node $n$ at slots $l_1$ and $l_2$ respectively such that $l_1<l_2$, then $len(C_2)-len(C_1)\geq \zeta \cdot (l_2-l_1)$. $\zeta$ is the speed coefficient.

    \item \emph{Chain quality}:  We say that the protocol satisfies chain quality with parameter $\upsilon$ if given a consecutive run of $l$ blocks on a chain $C$ adopted by an honest node, has at least $\upsilon \cdot l$ blocks generated by honest nodes. $\upsilon$ is the chain quality coefficient.


\end{itemize}

\subsection{Verifiable Random Function and Sybil resistance}
A Verifiable Random Function (VRF) is a pseudorandom function that computes a hash of an input value specific to a given private key. Along with this hash, it also produces a proof that allows the hash to be verified using the corresponding public key. It allows only the owner of the secret key to generate the output of a VRF, but anyone can verify it. VRFs have been used extensively in cryptocurrencies such as Algorand, Cardano and Ouroboros Praos. 

Sybil resistance is the property of a function to not give the adversary any advantage by splitting itself into multiple entities. VRFs maybe used to achieve Sybil resistance. Algorand uses it for cryptographic sortition, while \proto\ uses VRF for calculating block power.



\section{The Intuition behind \proto}
\label{ssec:op_oa}
We require that our protocol, \proto, satisfies the following two requirements:
\begin{itemize}
    \item To ensure security against FACs, the SBP must not be revealed before it publishes the block.
    \item Forks cost performance. Hence, forking amongst the honest nodes must be avoided. At any time, we must ideally have only one chain to be extended.
\end{itemize}

To fulfill the first requirement, the computation required by the BPSM must be done privately. For the second requirement, the BPSM must select exactly one block publisher as the SBP. Satisfying both of these requirements together is non-trivial, as we must privately, securely, provably, and efficiently select exactly one party in a multiparty weighted coin-toss with guaranteed output using a seed-randomness. The BPSM presented in this paper solves for both these requirements through the CSR. In contrast, v1 solves only the second requirement but not the first, while Praos solves only the first requirement but not the second.

Each node publishes its block in an attempt to have it selected as the next block added to the blockchain. To avoid forks, the honest nodes must reach to a consensus on a single chain among various possible ones. If nodes are able to together achieve this, they will be able to extend the agreed upon chain with a new block in the next slot. 
We introduce the metric \emph{block power} so nodes can evaluate competing blocks and reach at the same result. To ensure that our protocol is PoS-based, the block power must be dependent on the relative stake held by the publisher of this block.

Let us use an example to demonstrate the power of our metric based approach. Consider a fork, where nodes, $n_a$ and $n_b$, both are aware of two blocks, $B_1$ and $B_2$ viable for extension. Now, both Ouroboros and \btc, make no distinction between the two blocks. Hence, the node $n_a$, may extend block $B_1$, while node $n_b$, extends block $B_2$. This causes an extension of the fork. \proto\ gracefully avoids such problems by easily determining which block has a higher \emph{block power}. By embedding the metric of block power in every block, both nodes collectively decide to extend either block $B_1$ or block $B_2$, without ever going through an elaborate process to establish consensus or wait for one of the forks to be extended by the rest of the network.

\section{\proto\ Protocol}
\label{sec:ourproto}

First, we introduce the framework for our protocol, \proto. We then describe in detail the notion of block power and prove it is Sybil-resistant. Next, we describe various salient features of \proto\ such as block publisher selection mechanism (BPSM) and chain selection rule (CSR) (Section \ref{ssec:op_csr}). In Section \ref{ssec:op_bp} we explain how the block publisher builds and broadcasts the block. Lastly, we list the assumptions made while constructing the \proto\ protocol.
In Section \ref{ssec:op_sr}, we briefly discuss the security of \proto. 

\subsection{Framework}
\label{ssec:iop}

\proto\ builds on the Ouroboros framework to attain greatly improved performance while being resistant to a dynamic byzantine adversary with up to 50\% stake. We briefly describe the functionalities that we adopt from the Ouroboros protocols. To avoid redundancy, we do not discuss the details, nuances, and implementations of these functionalities, which are well presented in the Ouroboros protocol papers.

\subsubsection{Communication Network}
\begin{itemize}
    \item We consider a network, $N$, of honest nodes $\mathbb{H}$, having $r_h\geq1/2$ of the total relative stake. All nodes not in $N$, are considered to be adversarial who represent $r_a$ relative stake. $\mathbb{A}_a$ comprises of nodes with malicious intent, while $\mathbb{A}_h$ comprises of nodes that follow the protocol but are not in $N$ due to their high communication delay. The network is described in Fig. \ref{fig:netN}.
    \item The maximum communication delay (block propagation delay) between any two nodes during synchronous phase is at most $\tau$. In \proto, $t_{sl}$ is set equal to $\tau$. \\
\end{itemize}

\subsubsection{The Adversary}
We use a dynamic (\emph{mobile} \cite{yung2015mobile}) byzantine adversarial model (which in terms of ability, due to its dynamic nature, is more powerful than the one considered in v1, and the same as the ones in Praos and Algorand). The dynamic nature of the adversary enables \emph{Fully Adaptive Corruptions} (FACs). FAC is the ability of the adversary to corrupt any node instantaneously. FACs threaten the security of the protocol when the adversary can know that corrupting a certain portion of the relative stake is better than corrupting another portion of the relative stake (of the same size). However, the adversary is always bound by $r_h\geq1/2$.

In favor of the adversary, we assume that both $\mathbb{A}_a$ and $\mathbb{A}_h$ are part of the adversary. Thus, we define the adversary, $\mathbb{A}$, as $\mathbb{A}=\mathbb{S} \setminus \mathbb{H}$.
In our model, any $a\in\mathbb{A}$ can:
\begin{itemize}
    \item read all communication between all nodes instantly.
    \item show any chain that it is aware of, to any honest node.
    \item corrupt any node (turn any honest node into an adversarial node) at any given moment provided $r_h\geq1/2$.
    \item freely, privately, and instantly communicate amongst all its nodes. All the nodes in $\mathbb{A}$ are assumed to be united by a single objective, in favor of the adversary. Hence, the adversary is considered to be a single entity. \\
\end{itemize}

\subsubsection{Time and Slots}
\label{ssec:time-and-slots}
The protocol executes in a sequence of time periods called \emph{epochs} that are further divided into \emph{slots}. Each epoch contains a predefined number of slots. Each slot is of a predefined length (in units of time), $t_{sl}$ ($t_{sl}>0$). The slot number, $l$, (counted from the start of the blockchain execution, with the slot corresponding to the genesis block being slot $l=0$), uniquely identifies a slot, as well as the epoch $ep$ that the slot $l$ is a part of. 

\noindent\textbf{Comparing our Network Model with Ouroboros family and Algorand.} We use the same \emph{partial synchrony model} as Ouroboros Praos. Players have roughly synchronized clocks and can access the publicly available current time (using the \emph{Network Time Protocol} (NTP)) to determine the current slot. Algorand \cite{gilad2017algorand} can ensure security in bounded unknown delay based partially synchronous model, but it relies on strong synchrony assumptions to guarantee liveness of the protocol.

The difference in the local time of these players is assumed to be insignificant compared to the length of the slot. Ouroboros Chronos \cite{ouroboros_chronos} leverages proof of stake (PoS) to solve the global synchronization problem. We recommend using such mechanisms to remove the need of relying on a global clock. However, the local clocks of players are assumed to progress at approximately the same speed.
\\

\subsubsection{Pseudo-genesis block}

At the start of each epoch $ep$, there is a \emph{pseudo-genesis block} that enlists the stake distribution of all the stakeholders, as well as the seed randomness used to determine the SBP in this epoch.

$B_{PG}^{ep}=\{seed^{ep},\{\{pk_0,r_0^{ep}\},\{pk_1,r_1^{ep}\}, \ldots \}\}$. 

\begin{itemize}
    \item The stake distribution is as per the last block of the second last epoch, i.e., $ep-2$.
    \item The seed randomness is a result of a multiparty computation, \emph{private verifiable secret sharing} (PVSS) \cite{schoenmakers1999simple} (although in practice Cardano uses \emph{SCRAPE: SCalable Randomness Attested by Public Entities} \cite{cascudo2017scrape}). This seed randomness is based on the input of the SBP of the previous epoch, i.e., $ep-1$.
\end{itemize}

\begin{figure*}[!ht]
\begin{subfigure}{.4\columnwidth}
    \includegraphics[scale=0.2]{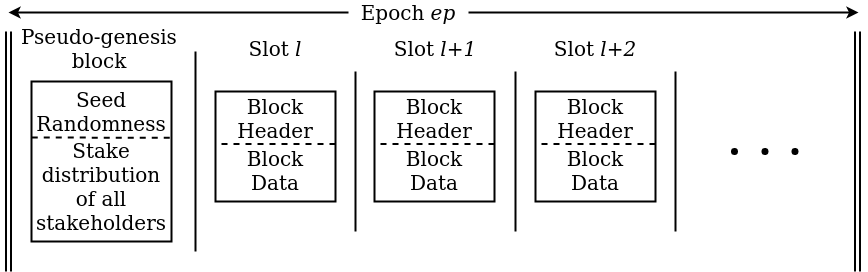}
  \caption{\centering Epochs and Slots in Ouroboros} \label{fig:es}
\end{subfigure}
  \begin{subfigure}{0.3\columnwidth}
      \includegraphics[scale=.14]{NetworkN.png}
  \caption{The Network $N$} \label{fig:netN}
  \end{subfigure}
  \begin{subfigure}{0.3\columnwidth}
      \includegraphics[scale=.25]{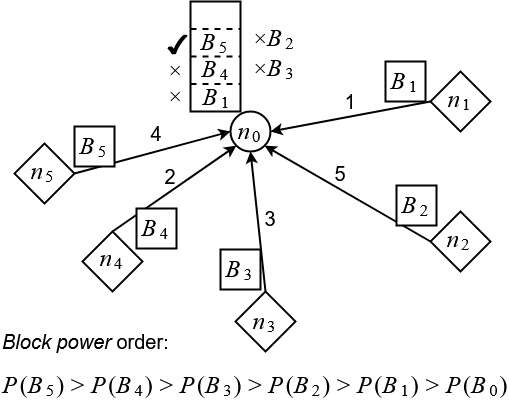}
  \caption{Broadcasting Blocks} \label{fig:comm}
    \end{subfigure}
    \caption{System Model Model}\label{fig}
\end{figure*}

As shown in Fig. \ref{fig:es}, this framework of epochs, slots, and pseudo-genesis blocks helps in restricting the power of the adversary.
The adversary should not know the resulting random seed while it can influence it. If this is not the case, the adversary can always simulate the most favorable way to influence the random seed. Also, the adversary should not be able to change its stake distribution amongst its nodes after observing the random seed. If this is not the case, then the adversary can distribute its stake optimally based on the random seed and exploit the BPSM to its advantage. \\

\subsubsection{Forward Secure Key Signatures}
We use \emph{forward secure key signatures} \cite{bellare1999forward,itkis2001forward,david2018ouroboros}, as used in Praos. Each node $i$ has a keypair consisting of \emph{public key} $pk_i$, and a \emph{master secret key} $msk_i$. The $pk_i$ does not change, while $msk_i$ is used to generate a temporary secret key $sk_i^l$ (starting from $l=0$), which can verifiably be used only for slot $l$ by node $i$, after which $sk_i^{l+1}$ is generated and $sk_i^l$ is \emph{irrecoverably} destroyed. 

When nodes publish a block in a certain slot, their achieved block power in that slot becomes public knowledge. The adversary thus knows exactly which nodes to corrupt. Forward secure key signatures prevent the adversary from corrupting an honest node to re-publish a block in its favor after the honest node has already published a block in that particular slot. \\

\subsubsection{Moving Checkpoints}
Similar to v1 and Praos, \emph{moving checkpoints} are used in \proto\ to prevent long-range attacks such as \emph{stake-bleeding attacks} \cite{gavzi2018stake}. \emph{Moving checkpoints} are states of the blockchain, appropriately far enough into the history, such that no honest nodes would disagree with them. These moving checkpoints are assumed to be reliably and convincingly communicated to all the honest nodes joining the execution of the protocol (or any other node that has not witnessed the evolution of the honestly maintained public blockchain to establish the checkpoint for itself).

\subsection{Block Power}

Ouroboros uses non-ideal VRFs, such as digital signatures, to stochastically select SBPs. However, to avoid manipulation of the selection probability through BPSM, an \emph{ideal-VRF} is required. This is achieved using the 2-Hash-DH \cite{jarecki2014round} construction, as presented in Praos.
 $$ VRF(sk_i^l,l,seed^{ep}) \rightarrow (\sigma_{uro}^{i,l,seed^{ep}},\sigma_{proof}^{i,l,seed^{ep}})\ $$
The above VRF takes as input:
\begin{itemize}
    \item the secret key of the node $i$ corresponding to slot $l$,
    \item the current slot number $l$,
    \item the seed randomness of epoch, $ep$
\end{itemize}
It produces a uniform random output of length $\kappa$ in bits. Any entity can verify the legitimacy of this output $\sigma_{uro}^{i,l,seed^{ep}}$, using $VRF_{VERIFY}(\sigma_{uro}^{i,l,seed^{ep}},\sigma_{proof}^{i,l,seed^{ep}},pk_i,l,seed^{ep})$. \\

Let us now define the numerical metric, \emph{block power}, using the following steps. Block power can be computed from the block header alone, and does not depend on the block data. For the optimal performance of \proto, we scale the relative stake, $r$, by a constant parameter $s$ to yield stake power $\alpha$, which is then used to determine the block power.

\begin{enumerate}
    \item $ \alpha_i^{ep} = r_i^{ep} \times s $ \\
    $r_i^{ep}$ is the relative stake of node $i$ in epoch $ep$. It is scaled up by $s$ ($s>0$), a constant scale factor defined in genesis block.
    \item $ \sigma_{nuro}= \frac{\sigma_{uro}}{2^{\kappa}} $ \\
    The uniform random output of the VRF defined above is normalized to lie in the range of $[0,1]$.
    \item $ P(B_C^l)= \sqrt[\alpha_i^{ep}]{\sigma_{nuro}} $ \\
    Now, we define the $\alpha_i^{ep}$ root of the normalized uniform output.
\end{enumerate}

The chain power, $P(C)$, of a chain, $C$, is calculated as the sum of block powers, $P(B)$, of all the blocks $B \in C$.

Our proposal for using block power solves the same purpose as \emph{Cryptographic Sortition} in Algorand. However, block power is much simpler and faster to calculate compared to the loop structure used in Cryptographic Sortition.

Please note: We overload the notation, $P(\cdot)$, to denote block power as well chain power; When a block (chain), $B_1$, has higher block (chain) power than another block (chain), $B_2$, we say that block (chain), $B_1$, is better than block (chain), $B_2$; We use the notation $B_C^l$, to refer to the block, $B$, that is a part of chain, $C$, that was published in slot $l$.

\subsection{Probability distribution function of Block Power}

We denote $W$ as the random variable for block power and $f_{\alpha} (w)$ is the \emph{pdf} of this block power. The normalized VRF output $\sigma_{nuro}$ is a uniform random variable.
Let $w$ be the power of some block $B$, corresponding to the VRF output $\sigma_{nuro}^{'}$.

\begin{align*}
    W &= \sqrt[\alpha]{\sigma_{nuro}} \\
    Pr( W \leq w) &= \int_{0}^{w} f_{\alpha} (w) \,dw \\
    Pr( W \leq w) &= Pr( \sqrt[\alpha]{\sigma_{nuro}} \leq \sqrt[\alpha]{\sigma_{nuro}^{'}} ) \\
    Pr( \sqrt[\alpha]{\sigma_{nuro}} \leq \sqrt[\alpha]{\sigma_{nuro}^{'}} ) &= Pr( \sigma_{nuro} \leq \sigma_{nuro}^{'} ) = \sigma_{nuro}^{'} \\
        \text{using above equations,}\ f_{\alpha} (w) &= \frac{d (\sigma_{nuro}^{'})}{dw} = \frac{d (w^{\alpha})}{dw}
\end{align*}

\begin{center}
    \boxed{f_{\alpha} (w) = {\alpha} w^{{\alpha}-1} } \label{eqn:sa_res}
\end{center}

Hence, we derive the probability distribution function of block power as a function of the block power $w$ and the stake power $\alpha$. \\

\subsubsection{Proving Sybil Resistance}

To be Sybil resistant, we must ensure that whether a node is represented as a single entity of stake $\alpha$ or two entities of stake power $\alpha_1$ and $\alpha_2$, the \emph{pdf} of the node's effective block power (the node's maximum block power, as one or more entities) remains the same. The probability of some node becoming the SBP should be same, regardless of division into smaller nodes or aggregation into bigger ones.

The probability that one of 2 nodes, $n_1$ and $n_2$, with scaled relative stakes $\alpha_1$ and $\alpha_2$, will get block power $w$ is equal to the sum: \\
 Pr($n_1$ has block power $w$ while $n_2$ has block power $\leq$ $w$) + Pr($n_2$ has block power $w$ while $n_1$ has block power $\leq$ $w$). 

Upon computing this sum, we find that it equals the probability to get block power $w$ by a node $n$ with scaled relative stake $\alpha=\alpha_1+\alpha_2$. 
The derivation has been skipped due to space constraints.

\begin{center}
    \boxed{$$f_{\alpha}(w)= f_{\alpha_1}(w)\int_0^w f_{\alpha_2}(y) dy + f_{\alpha_2}(w)\int_0^w f_{\alpha_1}(y) dy$$}
\end{center}

Since the \emph{pdf} in Eqn. \ref{eqn:sa_res} satisifies the above equation for all $\alpha_1,\alpha_2>0$, the proposed block power is resilient to Sybil attacks. 

\subsection{BPSM and CSR}
\label{ssec:op_csr}

We propose the following \emph{Chain Selection Rule} (CSR); to be followed by node, $i$, in slot, $l$. The chain $C_{csr}$ is picked from a known set of chains known and extended by publishing a block in slot $l$. To ensure that the execution of the protocol w.r.t. slots is respected, only chains of a certain (for a given slot) length are considered valid. 
\begin{enumerate}
    \item From the set of chains, $S_{{view}(i,l)}^{Chains}$, select a subset of chains $S_{{view}(i,l)}^{validChains}$, such that $\forall C \in S_{{view}(i,l)}^{validChains}, len(C)=l-1$.
    \item Calculate the chain power $P(C)$, of every chain $C$ in $S_{{view}(i,l)}^{validChains}$. The chain with the maximum chain power is the $C_{csr}$. 
    \item Publish a block extending chain $C_{csr}$.
\end{enumerate}

Our construction of block power ensures that the probability of two chains having the same chain power is very low. In the event that this does happen, the next SBP will extend one of them, all nodes in $N$ will then accept this extension. 
At the start of each slot, every node publishes a block, extending the chain selected by the CSR. Since all nodes publish blocks and the block that is to be extended upon is determined by the CSR, the SBP for a slot can be said to be determined by the CSR in the next slot. Hence, the \emph{Block Publisher Selection Mechanism} (BPSM) depends on the CSR. 

\subsection{Block Publishing}
\label{ssec:op_bp}
Block publishing comprises of building a block and then broadcasting it. To publish a block in slot, $l$, a node, $i$, must build the block data, $Bd_i^l$, and the block header, $Bh_i^l$, as prescribed below. The block $B_i^l = \{Bh_i^l,Bd_i^l\}$, is then broadcast to the network. In \proto, to optimize the communication process, only the best block seen is propagated forward by the nodes instead of their own block.

\paragraph{Block building}
Block building is done at the start of the slot $l$, after the chain $C_{csr}$ has been selected according to CSR as the one to be extended. The node $i$, collects transactions, $\{tx_0, tx_1, \ldots \}$, received by the end of slot $l-1$, that have not yet been added to the blockchain and forms the block data, $Bd_i^l=\{tx_0, tx_1, \ldots \}$. The block header is then built in the format, $Bh_i^l=\{pk_i,r_i^{ep},l,\{hash(B_{C_{csr}}^{l-1}),null(B_{C_{csr}}^{l-1})\},$ $\{\sigma_{uro}^{sk_i^l,l,seed^{ep}},\sigma_{proof}^{sk_i^l,l,seed^{ep}}\},MTR(Bd_i^l)\}$; where $i$ is the node with public key, $pk_i$, and relative stake, $r_i^{ep}$; $l$ is the slot number; $ep$ is the epoch number; $\{\sigma_{uro}^{sk_i^l,l,seed^{ep}},\sigma_{proof}^{sk_i^l,l,seed^{ep}}\}$ is generated by $VRF(sk_i^l,l,seed^{ep})$; $MTR(Bd_i^l)$ is the \emph{Merkle tree root} of the block data $Bd_i^l$, it is what binds the block header to the block data $Bd_i^l$ (also, fixes order of $tx$ in $Bd_i^l$); $\{hash(B_{C_{csr}}^{l-1}),null(B_{C_{csr}}^{l-1})\}$ binds block $B_i^l$ as extending the chain ${C_{csr}}$ beyond block $B_{C_{csr}}^{l-1}$, and denotes whether the block $B_{C_{csr}}^{l-1}$, was a null block. 

The block, $B_i^l$, is now ready to be broadcast.
Note, the block header is sent along with a digital signature over itself for sender authentication.

\paragraph{Broadcasting the best block}
Since all nodes will have valid blocks that are contenders for the best block, waiting to collect all blocks from all users in $N$ will require $\tau$ to be very high. Instead, the nodes only download and broadcast the block data of the best block header seen yet.
From the perspective of block propagation time, this method of communication is equivalent to communicating a single valid block.
Since the block header is sufficient to determine the block power of a given block, a node is aware of a block and its power as soon as it receives its block header. Once a node is aware of a block, it may or may not attempt to download its block data.
An honest node should always try to propagate the best block header it is aware of, to all other nodes, at any given moment.

Consider a network, as shown in Fig. \ref{fig:comm}. W.L.O.G., let us say that the node $n_0$ is the one that wants to build the next block and hence needs to know which is the best block in the current slot. The nodes $n_0, n_1, n_2, n_3, n_4, n_5$ have published blocks, $B_j=\{Bh_j,Bd_j\}; j=\{0..5\}$ with $P(B_1)<P(B_2)<P(B_3)<P(B_4)<P(B_5)$. Say, node $n_0$, sees the blocks in the following order: $B_1$, $B_4$, $B_3$, $B_5$, $B_2$. Now, first, $n_0$ sees $Bh_1$, and since it is better than $Bh_0$, and hence the best block it is aware of at the moment, begins to download $Bd_1$. Whether or not $n_0$ has finished downloading $Bd_1$, once it sees $Bh_4$, it will stop downloading $Bd_1$ and start downloading $Bd_4$. $n_0$, will, however, disregard $B_3$, as it already is aware of a better block, i.e., $B_4$. However, when $n_0$ hears of $Bh_5$, it will switch to downloading $Bd_5$ instead, as its block power is higher. Again, it will disregard $B_2$, as it is aware of a better block $B_5$.

Although the above disussion pertains to broadcasting blocks, the nodes, in fact, broadcast chains, in essentially the same manner as discussed above. This broadcasting of chains devolves to broadcasting blocks when the chains differ only by one block, i.e., the one published in slot $l$. We omit the discussion on broadcasting chains, i.e., the ones that differ from $C_{csr}$, by more than one block for simplicity and tractability. However, it is important to note that the nodes always propagate the best chain seen so far (similar to the best block propagation as discussed above).

\subsubsection{Block Confirmation}
\label{ssec:op_bc}
The nodes confirm all, but the most recent $k$ blocks, of the chain selected by the $C_{csr}$. This confirmation is consistent with \emph{k-finality} of \proto.

\subsection{Assumptions}
\label{ssec:assumptions}
We make the following assumptions:
    \begin{itemize}
        \item Ouroboros do not explicitly state the block size or the number of transactions per block. So, for a fair comparison, we assume that $tpb$ (typically we assume $tpb=2000$) in \proto\ and Ouroboros are the same as Bitcoin.
        \item Given the current state of the internet, it is safe to say that a typical block, similar to Bitcoin's, consisting of 2000 transactions, reaches 95\% percent of the nodes in 40 sec. \cite{bitcoin_network_monitor-dsn_research_group,blockchain.com_tpb} (\cite{decker2013information} does not discuss the average block size or the number of transactions per block). For ease of analysis, we ignore the 5\% tail and say that within 40 sec, all nodes hear of the block, i.e., that 40 sec is our upper bound on propagation, i.e., $\tau=40 $ sec.\footnote{We believe $\tau$ can be much less than $40$ sec as the advent of technology. The lower the $\tau$, better the security (w.r.t. a given $t_{sl}$).}
        \item As is typical in the analysis of blockchain protocols, we assume that \emph{honest behavior} of the honest nodes is motivated and ensured by the reward schemes. 
        \item In our analysis of $tps$, for the ease of comparison, we assume that the adversary and all active honest nodes publish blocks with their block data at every opportunity they get. In light of this assumption, we can claim that, $tps=\frac{tpb \times \zeta}{t_{sl}}$.
    \end{itemize}

\begin{figure}[!tbp]
  \centering
  \begin{minipage}[b]{\columnwidth}
\begin{center}
      \includegraphics[scale=.2]{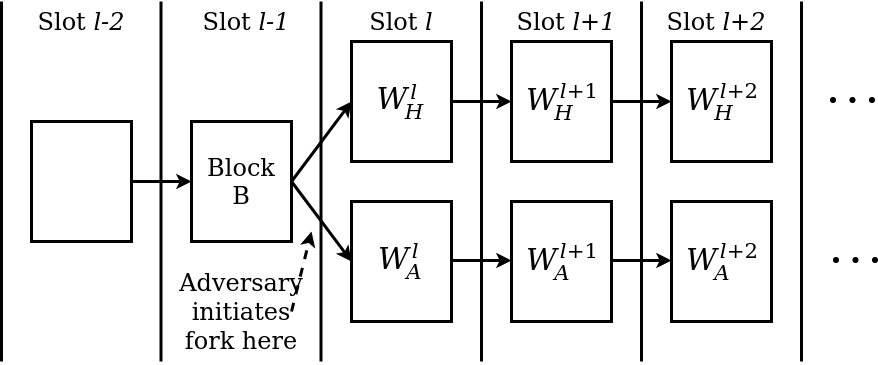}
\caption{The forking attempt by the adversary to violate finality} \label{fig:CP}
\end{center}

  \end{minipage}
\end{figure}

\begin{figure}
\caption{\proto\ Protocol}
\label{fig:proto}

\begin{minipage}{\columnwidth}
\begin{small}
\begin{algorithm}[H]
\renewcommand{\thealgorithm}{}
\floatname{algorithm}{\proto\ \emph{Protocol Pseudo-code};}

\caption{}
\label{protocol1}
\begin{flushleft}
Followed by node $i$ in slot $l$:

INPUT: \{$sk_i^l$, $r_i^{ep}$, $seed^{ep}$, $S_{{view}(i,l)}^{Chains}$, $s$, $\{tx_0, tx_1, \ldots \}$\}

Step 1: \emph{Chain selection}
\end{flushleft}
\begin{algorithmic}[1]
\STATE From $S_{{view}(i,l)}^{Chains}$, select the subset $S_{{view}(i,l)}^{validChains} : \forall C \in S_{{view}(i,l)}^{validChains} | len(C) = l - 1 $.
\STATE $\forall C \in S_{{view}(i,l)}^{validChains}$, calculate, $P(C)$.
\STATE Select the chain,

$C_{csr} : C_{csr} = \argmax_{C \in S_{{view}(i,l)}^{validChains}} P(C)$
\end{algorithmic}

\begin{flushleft}
Step 2: \emph{Block publishing}
\end{flushleft}
\begin{flushleft}
\qquad a) \emph{Block building}
\end{flushleft}
\begin{algorithmic}[1]
\STATE Build $Bd_i^l=\{tx_0, tx_1, \ldots \}$, and obtain $MTR(Bd_i^l)$.
\STATE $\{\sigma_{uro}^{i,l,seed^{ep}},\sigma_{proof}^{i,l,seed^{ep}}\} \leftarrow VRF(sk_i^l,l,seed^{ep}) $
\STATE Obtain $\{hash(B_{C_{csr}}^{l-1}),null(B_{C_{csr}}^{l-1})\}$
\STATE Build $Bh_i^l=\{pk_i,r_i^{ep},l,\{hash(B_{C_{csr}}^{l-1}),null(B_{C_{csr}}^{l-1})\}, \allowbreak \{\sigma_{uro}^{sk_i^l,l,seed^{ep}},\sigma_{proof}^{sk_i^l,l,seed^{ep}}\},MTR(Bd_i^l)\}$
\end{algorithmic}

\begin{flushleft}
\qquad b) \emph{Block broadcasting}
\end{flushleft}
\begin{algorithmic}[1]
\STATE Set $C_{broadcast}=\{C_{csr},B_i^l\}$
\WHILE{Current Slot $== l$}
\STATE Listen and receive $C_{rec}$ from other nodes
\IF {$P(C_{rec})>P(C_{broadcast})$}
\STATE Set $C_{broadcast}=C_{rec}$
\ENDIF
\STATE Broadcast $C_{broadcast}$ 
\ENDWHILE
\end{algorithmic}

\end{algorithm}

\end{small}

\end{minipage}
\hfill
\begin{minipage}{\columnwidth}
\begin{small}
\end{small}
\end{minipage}
\end{figure}

We conclude the presentation of \proto\ with its pseudo-code, given in Fig. \ref{fig:proto}. The essence of our protocol can be summarized as follows: The nodes collectively know which chain to adopt, which then becomes the honest chain. As soon as a fork is witnessed, adversarial or not, the nodes know whether or not to adopt it; the nodes are not confused, as they would be in protocols that do not differentiate between published blocks, such as \btc\ or Ouroboros.
However, the adversary can still attempt to fork and develop chains and use them to violate finality. In the next section, we briefly discuss security and prove that \proto\ is secure against such attempts. 

\subsection{\proto\ Security and Performance}
\label{ssec:op_sr}
\subsubsection{\proto\ Security Analysis}
We can formally prove the security of \proto as we to establish the three prerequisite properties (Section \ref{ssec:reqbp}) of a blockchain protocol in Theorem \ref{theo:sr}. For lack of space and ease of presentation, we skip the proofs.

\begin{restatable}[]{theorem}{theosr}
\label{theo:sr}
    The probability that any of; common prefix with parameter $k$, chain growth with parameter $\zeta=1$, chain quality with parameter $\upsilon=1/k$ are violated in the lifetime of the protocol, thereby violating liveness and persistence is:
        $$\varepsilon_{lp} \leq 2\varepsilon_{cp}$$
\end{restatable}


\begin{restatable}[]{proposition}{propfac}
\label{prop:fac}
    \proto\ is resilient to \emph{Fully Adaptive Corruptions} (FACs).
\end{restatable}


\proto\ does not require a \emph{Forkable Strings} analysis, as presented in v1 and Praos. Due to the elegance and naturality of our approach, \proto\ avoids the inherent complexities of the Ouroboros protocols.
All observed (public) forks are immediately and trivially resolved using chain power as the nodes will only build on the chain with the highest chain power. If an adversarial chain with higher chain power than an honest chain is revealed before it can violate finality, that adversarial chain becomes the new honest chain. To violate finality (and hence common prefix), an adversarial chain must have higher chain power than the honest chain while forked for at least $k$ blocks; we prove that the probability of this is exponentially bounded with $k$. \\

\subsubsection{\proto\ Performance Analysis}
We optimize \proto\ for improved performance, but do not mention them here for sake of simplicity and lack of space. These details may be provided if requried.
Using our claimed performance, we present a comparison of $t_f$ and $tps$, between \btc, Ouroboros v1, and \proto, in Tables \ref{tab:com_f} and \ref{tab:com_tps}.
The summary of our analysis is, \proto\ improves $t_f$ by factor of 3 and marginally improves  $tps$  over  Ouroboros v1. 
\begin{table}[!th]
\caption{Comparison of $tps$} \label{tab:com_tps}
\begin{center}
 \begin{tabular}{|c|c|} 
 \hline
 & $tps$\\ [0.5ex]
 \hline
 Bitcoin & $\frac{tpb}{avg. time per block (sec)}=\frac{2000}{600}=3.3$   \\
 \hline
Ouroboros v1 & $\frac{tpb \times r^{active}}{t_{sl}} \implies tps<50$ $\forall$ $r^{active}<1$     \\
 \hline
 \proto & $\frac{tpb}{t_{sl}} = 50$ $\forall$ $r^{active}>0$\\
 \hline
\end{tabular}
\end{center}
\end{table}
\begin{table}[!t]
\caption{Comparison of Time to Finality (in minutes)}
\subcaption*{BTC: Bitcoin, v1: Ouroboros v1, QS: \proto}\label{tab:com_f}
\begin{center}
\resizebox{0.5\columnwidth}{!}{
 \begin{tabular}{||*{7}{c|}|} 
 \hline\
 \diagbox[width=1.5cm, height=.7cm]{$r_a$}{$1-\eta$} & \multicolumn{3}{c|}{0.99} & \multicolumn{3}{c||}{0.999} \\ [0.5ex]
 & BTC & v1 & QS & BTC & v1 & QS \\
 \hline\hline
 0.10 
 &40 &6 &2  
 &50 &10 &4    \\
 0.15 
 &40 &10 &4   
 &80 &16 &6    \\
 0.20 
 &70 &14 &5  
&110 &24 &8  \\
 0.40 
 &580 &183 &55 
 &890 &296 &90    \\
 0.45
 &2200 &831 &226  
 &3400 &1327 &361    \\
 0.47 
 &5970 &2506 &632 
 &8330 &3969 &1041    \\
 0.48
 &- &5991 &1434  
 &- &9438 &2335    \\
 \hline
\end{tabular}
}
\end{center}

\end{table}

\section{Discussion on \proto\ Parameters}
\label{sec:disc}

In this section, we discuss how to set the three parameters relevant to the protocol \proto. Two parameters are endogenous to the protocol, namely, $s$ and $t_{sl}$. 
\begin{itemize}
    \item $s$-- the factor by which the relative stake of a node is scaled. In \proto, $s$ affects the $t_f$  significantly and hence should be chosen to be optimal. We recommend using $s=8$, although any $s>4$ should suffice. 
    
    \item $t_{sl}$ is the time slot length in units of time. $t_{sl}$ greatly impacts transaction per second ($tps$), time to finality ($t_f$), and relative stake of the honest nodes ($r_h$). Ideally, $t_{sl}\geq \tau$ where $\tau$ is the upper bound on the block propagation delay. Decreasing $t_{sl}$, increases $tps$ and reduces $t_f$ which is very favourable. However, this comes at the risk of reducing it below the actual value of $\tau$, causing $r_h$ to drop below half, compromising the security of the protocol. Given the current state of the internet we suggest $t_{sl}=\tau=40$ sec (from \cite{bitcoin_network_monitor-dsn_research_group,decker2013information}). Once we set $t_{sl}$, it need not be changed as internet connectivity will only get better.
    
    \item Another crucial parameter is $k$, which is exogenous to the protocol. It is an important factor from the perspective of the consumer of the blockchain. $k$ is the number of blocks that a node should wait to confirm a block for a given confidence level $1-\eta$ against an adversary with relative stake $r_a$. As the block gets deeper into the blockchain, the confidence of the block's placement in the ledger increases. The longer the asynchronous period (during which network is controlled by a \emph{FAC} adversary), the more time is needed for finality while ensuring security.

\end{itemize}

\section{Related Work}
\label{sec:rel}

 \paragraph*{Bitcoin Protocol}
Bitcoin's popularity has attracted much research on this subject. Analysis of the functioning of the Bitcoin protocol, in \cite{garay2015bitcoin,pass2017analysis}, provides a fundamental understanding of how it implements a robust transaction ledger, while a study of Bitcoin from a game-theoretic perspective, such as in \cite{lewenberg2015bitcoin,kwon2019bitcoin, Badertscher2018}, provides insight into the realistic operation. The protocol vulnerabilities are well explored in works such as \cite{eyal2018majority,carlsten2016instability,zhang2019lay,apostolaki2017hijacking, Badertscher2021}. There have been several efforts to improve the Bitcoin protocol, such as the one discussed in \cite{sompolinsky2013accelerating}. There is also much research based on utilizing blockchain technology for different purposes, as discussed in \cite{dziembowski2019perun,tomescu2017catena,mauw2018distance}.
 \paragraph*{PoS-based Protocols}
 As pointed out in the Introduction, to maintain Bitcoin, the nodes consume a large amount of power due to PoW-based BPSM. To avoid this massive power consumption, researchers explored alternative modes of mining currencies. PoS has been one of the most popular such modes.
 There have been several approaches to PoS-based protocols apart from the Ouroboros protocol, such as; \emph{PPcoin} \cite{king2012ppcoin}, which uses coin age along with PoS and PoW; Ethereum's \emph{Casper} protocol \cite{buterin2017casper}, that uses validators that lose their stake when they behave maliciously; \emph{Snow White} protocol \cite{bentov2016snow}, that too uses epochs and is provably secure; \emph{Algorand} protocol \cite{gilad2017algorand}, that uses \emph{Byzantine Fault Tolerant} mechanisms to achieve consensus and is also provably secure, though it requires $\frac{2}{3}$ majority of honest nodes. Other approaches include Dfinity~\cite{Abraham2018DfinityCE} which combines notarisation system, VRF and PoS  and propose new consensus algorithm. This protocol however, is not secure against adaptive corruptions. 

 Apart from independent protocols, there has been significant research on concepts that could potentially improve or augment PoS protocols. Ouroboros Crypsinous \cite{kerber2018ouroboros} already utilizes one such technology, \emph{Zerocash} \cite{sasson2014zerocash}, to provide privacy. PoS protocols could also use \emph{context-sensitive transactions} \cite{larimer2013transactions,larimer2018delegated} as discussed in \cite{gavzi2018stake} to do without moving checkpoints. Efforts such as \emph{Scalable Bias-Resistant Distributed Randomness} \cite{syta2017scalable}, could aid in improving the randomness beacons used in PoS protocols. Concepts such as, sharding as demonstrated in \emph{Omni-Ledger} \cite{kokoris2018omniledger}, and \emph{Proof-of-Stake Sidechains} \cite{gazi2019proof}, could potentially be applied to scale the throughput of blockchains.


\section{Future work and Conclusion}
\label{sec:conc}
PoS-based protocols are attracting attention in the literature but may potentially be vulnerable to FACs, as in v1. In this paper, we proposed a novel PoS-based blockchain protocol, namely, \proto. To design it, we introduced a block power metric resistant to Sybil attacks (Eqn \ref{eqn:sa_res}). We showed that \proto\ satisfies chain prefix, chain growth, and chain quality properties with appropriate parameters (Theorem \ref{theo:sr}). We also showed that \proto\ is resistant to FACs (Proposition \ref{prop:fac}). 
Our analysis showed that \proto\ performs better than the Ouroboros protocols for $t_f$ (time to finality; by a factor of 3) and $tps$ (transaction per second). 
We leave it for future work to explore the application of our BPSM and CSR to build blockchain protocols outside the Ouroboros framework.

We believe the concept of \emph{block power} and {chain power} metrics help design PoS-based blockchain protocols. Possible applications of Sybil attack-resistant functions in other scenarios involving Sybil attacks need to be explored further.

\section{Acknowledgement}

The research was partially funded through National Unified Blockchain Framework by MEITY, India and partially by Ripple IIIT Hyderabad CoE for Blockchain.

\newpage


\begin{appendix}

\section{Security Analysis}
\label{sec:sa}

\begin{claim}
\label{claim:bi}
Let $X_1, X_2,\ldots, X_M$ be independent random variables with zero mean, such that $|X_l| \leq K \: \forall \: l \in [M], K>0$. Also, $E[{X_l}^2]\leq \sigma_l^2$. Then, $\forall \lambda>0$:
$$\Sigma_{M=k}^\infty Pr \left(\frac{1}{M}\Sigma_{l=1}^M X_l \geq \lambda \right) \leq \frac{e^{- c\cdot k}}{1 - e^{-c}}$$
where, $c=\frac{\lambda^2}{2(\sigma_l^2+K\lambda/3)}$

\end{claim}
\emph{Proof.} Using \emph{Bernstein's inequality} \cite{bernstein} for $X_1, X_2,\ldots, X_M$,:
 $ \forall\lambda>0$,
\begin{align*}
Pr \left(\frac{1}{M}\Sigma_{l=1}^M X_l \geq \lambda \right) & \leq \exp(\frac{-M\lambda^2/2}{\sigma_l^2+K\lambda/3}) \\  
\implies \Sigma_{M=k}^\infty Pr \left(\frac{1}{M}\Sigma_{l=1}^M X_l \geq \lambda \right) & \leq \frac{e^{- c\cdot k}}{1 - e^{-c}}
\end{align*}
where, $c=\frac{\lambda^2}{2(\sigma_l^2+K\lambda/3)}$ \qed

\begin{lemma}
    The probability that \proto\ does not satisfy the common prefix property with parameter $k$ is given by:
    $$\varepsilon_{cp} \leq \frac{Le^{- ck}}{1 - e^{-c}} \text{ where, $c=\frac{\lambda^2}{2(\sigma_l^2+K\lambda/3)}$}$$

\end{lemma}

\emph{Proof.} Let $E_1$ be the event when the adversary holds a chain that forks by more than $k$ blocks and has higher chain power w.r.t. the honest chain, in the lifetime of the protocol. When this event occurs, the adversary can show its chain to the honest nodes, making them accept it. When the honest nodes accept a chain that forks from their own by more than $k$ blocks, they are forced to change a confirmed block. This violates finality, and in turn, violates common prefix.
To violate finality, w.r.t. a particular block $B'$, the adversary must fork the chain at any point before the block $B'$, say at block $B$ and then have a better chain after at least $k$ blocks from $B$. Let $E_1^B$ denote the event that finality is violated with a fork starting at a certain block $B$.

Now we establish the probability, $\eta=Pr(E_1^B)$, that the adversary can violate finality with a fork starting at block $B$. Since this fork can start at any point in the lifetime of the protocol, we can say that the probability, $\varepsilon_{cp}=Pr(E_1)$, that common prefix is ever violated is at most $L \times \eta$, where $L$ is the lifetime of the protocol in slots.

To calculate the upper bound on $\eta$, we use Bernstein's inequality. Using this, we bound the mean of the power of the blocks from a given block, say $B$. If at any point, after $k$ blocks from block $B$, the mean power of the adversarial chain exceeds that of the honest chain then, we say event $E_1^B$ occurs, violating finality.

To use Claim \ref{claim:bi}, we take:
\begin{itemize}
    \item $W_A^l=\alpha_A x^{\alpha_A-1}$ and $W_H^l=\alpha_H x^{\alpha_H-1}$ to be the random variables, representing the power of the blocks of the adversary and the honest nodes respectively, in slot $l$. Here, $\alpha_A = s \times r_a$ and $\alpha_H = s \times
    r_h$. Please refer to Fig. \ref{fig:CP}.
    \item $X_l = W_A - W_H + (E(W_H)-E(W_A))$.
    \item $\lambda=E(W_H)-E(W_A)=\frac{\alpha_H}{\alpha_H+1} - \frac{\alpha_A}{\alpha_A+1}>0$, representing the advantage of honest nodes.
    \item $K = 1+\left(\frac{\alpha_H}{\alpha_H+1} - \frac{\alpha_A}{\alpha_A+1}\right)$
    \item $\sigma_l^2 =  \left(\frac{\alpha_A}{\alpha_A+2}-\left({\frac{\alpha_A}{\alpha_A+1}} \right)^2 \right)+\left(\frac{\alpha_H}{\alpha_H+2}- \left({\frac{\alpha_H}{\alpha_H+1}} \right)^2 \right)$
\end{itemize}

Thus:
\begin{itemize}
    \item $E(X_l)=0$
    \item $\begin{aligned}[t]
            |X_l| &\leq1+(E(W_H)-E(W_A)) \\
                  &= 1+(\alpha_H/(\alpha_H+1) - \alpha_A/(\alpha_A+1)) = K
            \end{aligned}$
    \item $\begin{aligned}[t]
            E[X_l]^2 & =\sigma^2(W_A - W_H + (E(W_H)-E(W_A)))\\
            &= \sigma^2(W_A) + \sigma^2(W_H) \qquad \text{, since } E(X_l)=0\\
            & = \left(\frac{\alpha_A}{\alpha_A+2}-\left({\frac{\alpha_A}{\alpha_A+1}} \right)^2 \right)\\
            & \qquad \qquad \qquad + \left(\frac{\alpha_H}{\alpha_H+2}- \left({\frac{\alpha_H}{\alpha_H+1}} \right)^2 \right)
            \end{aligned}$
\end{itemize}

Therefore by Claim \ref{claim:bi},
$\eta \leq \frac{e^{- ck}}{1 - e^{-c}}$, 
where, $c=\frac{\lambda^2}{2(\sigma_l^2+K\lambda/3)}$

Hence, the probability that common prefix is violated is,
$\varepsilon_{cp} \leq \frac{Le^{- ck}}{1 - e^{-c}}$, 
where, $c=\frac{\lambda^2}{2(\sigma_l^2+K\lambda/3)}$

Thus common prefix property is established.\qed

\begin{figure*}
\begin{subfigure}{0.49\textwidth}
\centering
\includegraphics[scale=0.60]{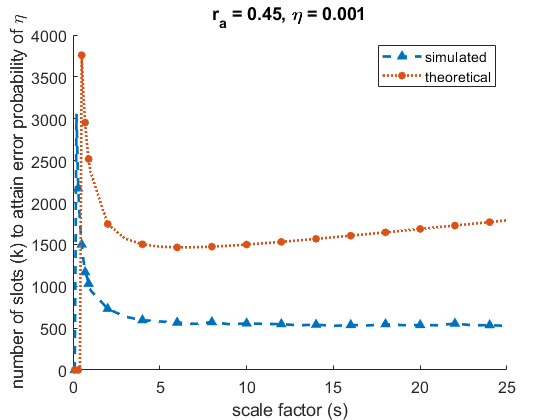}
\caption{Finality as a function of $s$} \label{fig:s_sim}
\end{subfigure}
\begin{subfigure}{0.49\textwidth}
\centering
\includegraphics[scale=0.60]{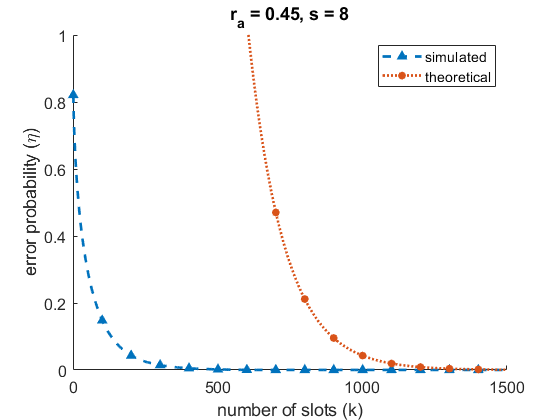}
\caption{Error probability $\eta$ as a function of $k$} \label{fig:eta_k}
\end{subfigure}
\caption{Simulation and Theoretical Analysis}
\end{figure*}

As shown in Fig. \ref{fig:s_sim} the plot of practical $t_f$ vs $s$ resembles an elbow curve. That is because as $s$ increases, even though the difference between the expectation of the block power of the honest nodes and the block power of the adversary decreases, the variance, too, decreases. This effect tapers out after $s=8$. Thus, we choose $s=8$ for the best practical finality.

Refer to Fig. \ref{fig:eta_k} for the error probability (violation of finality) $\eta$ as a function of $k$.

\begin{proposition}
\label{prop:bes}
    In \proto, a block is added to the chain in each slot, provided $r^{active}>0$.
\end{proposition}
\emph{Proof.} In \proto, the BPSM depends on the CSR. The CSR used in \proto\ assigns a chain power for each valid chain and selects the one with the highest power, and hence will be able to select a chain as long as there is at least one valid chain. To have a valid chain for the consideration by the CSR in the current slot, even one block is sufficient to have been published in the previous slot. Hence, even if there is just one active node in a given slot, there will be a block added to the chain in that slot.

\begin{lemma}
    \proto\ satisfies the chain growth property with parameter $\zeta=1$, given that $r^{active}>0$.
\end{lemma}
\emph{Proof.} Let $E_2$ be the event where there is no block added to the chain in a slot in the lifetime of the protocol. To satisfy chain growth with parameter $\zeta=1$, a block must be added to the chain in each slot, i.e., $Pr(E_2)=0$. For this, the BPSM must select a node in each slot.

However, by Proposition \ref{prop:bes}, $Pr(E_2)=0$ if $r^{active}>0$ for \proto. Therefore \proto\ satisfies the chain growth property with parameter $\zeta=1$, provided $r^{active}>0$.
Thus, the chain growth property is established. \qed

\begin{lemma}
    The probability that \proto\ does not satisfy the chain quality property with parameter $\upsilon=1/k$ is $\varepsilon_{cp}$.
\end{lemma}
\emph{Proof.} Let $E_3$ be the event that there is a run of $k$ or more consecutive blocks, published by the adversary, on the chain, $C$, held by an honest node, in the lifetime of the protocol.
Let $E'_3$, be the event where the sum of power of $k$ consecutive adversarial blocks is higher than the sum of power of the best corresponding honest blocks, in the lifetime of the protocol. Clearly, $E_1$ subsumes $E'_3$ (since if $E'_3$ occurs; $E_1$ too must occur). Now, for $E_3$ to occur, $E'_3$ must occur. Thus, $Pr(E_1)\geq Pr(E_3)$.

Now, evidently, when $E_3$ does not occur, there is at least one honest block in every run of $k$ consecutive blocks. Hence, when $E_3$ does not occur, $\upsilon>=1/k$. Therefore, the probability that chain quality with parameter $\upsilon=1/k$ is violated is at most $\varepsilon_{cp}$.

Thus, the chain quality property is established.\qed

\theosr*
\emph{Proof.} The above theorem follows from Lemmas 1, 2, 3 and the union bound $Pr(E_1 \cup E_3)$.\qed

\propfac*
\emph{Proof.} For a protocol to be resilient to FACs, it must be resilient against attempts of the following two corruptions:
\begin{itemize}
    \item \emph{Posterior corruption}: It is the adversary's ability to corrupt a node, use its stake and hence its block power, to publish a block in a past slot (w.r.t. the current slot).
    \item \emph{Anterior corruption}: It is the adversary's ability to corrupt a node that \emph{might} have the best block in the future (w.r.t. the current slot).
\end{itemize}

\proto\ avoids posterior corruption by using forward secure key signatures, as in Praos. Although anterior corruption by itself is not an issue, if the adversary knows which nodes \emph{will} have the best block with certainty or disproportionately high probability, then the adversary will corrupt those nodes and gain an advantage over the honest nodes. To avoid this, the adversary must not know which node will have the best block in the future. To this end, the nodes that attempt to publish a block must not reveal their block power until they publish the block. This is exactly the approach in \proto. In \proto, the block power is given in the block header, and the block header is not revealed until the block is published. Hence, the block power is not revealed until the block is published.

Thus, we say that \proto\ is resilient to FACs.\qed

Please refer to Section \ref{sec:sa_aa} for the analysis of attack strategies against \proto.

\section{Analysis of Attack Strategies against \proto}
\label{sec:sa_aa}

In this section, we discuss possible attack strategies and show that they are futile against \proto.

\paragraph{Sybil attack} Consider two scenarios: i) in which there is a node with stake power $\alpha_0$ and ii) there are two nodes with stake power $\alpha_1$ and $\alpha_2$. In the first scenario, there will be one value of block power in consideration, whereas, in the second scenario, there are two values. However, in the second scenario, only the higher block power is relevant as the CSR will (and hence the BPSM) select the block with the maximum block power. To avoid a Sybil attack, we need that in both the scenarios, the \emph{pdf} of the relevant block power to be the same; else the adversary has an incentive to split or aggregate stake power to have a higher probability of getting selected as SBP. To this end we must ensure that the \emph{pdf} of the block power in the first scenario must equivalent to the \emph{pdf} of the maximum of the two block powers in the second scenario, i.e., we need a block power function that satisfies: $$f_{\alpha_0}(x)= f_{\alpha_1}(x)\int_0^x f_{\alpha_2}(y) dy + f_{\alpha_2}(x)\int_0^x f_{\alpha_1}(y) dy$$
    where $\alpha_0=\alpha_1+\alpha_2 \; \forall \; \alpha_1,\alpha_2>0$, which is satisfied by $f_{\alpha}(x)=\alpha x^{\alpha-1}$.
    
    Also the probability that a node, with stake $\alpha_1$, wins over a node, with stake $\alpha_2$, is: $$\int_0^1f_{\alpha_1}(x)\int_0^x f_{\alpha_2}(y) dy dx=\frac{\alpha_1}{\alpha_1 + \alpha_2}$$

\paragraph{Double spending attack} In this sort of an attack, the adversary attempts to replace a certain block $B'$ on the chain of an honest node after it has confirmed the block $B'$. To do this, the adversary must show a better chain that forks by at least $k$ blocks, starting from before block $B'$, to an honest node that has confirmed block $B'$. This attack is ineffective, given that the common prefix property is established. However, the adversary can attempt to reduce the effectiveness of the stake power of $N$ or increase the effectiveness of its own stake power. It can attempt to do so in the following two ways, neither of which are effective:
\begin{itemize}
    \item \emph{Split $N$ attack}: Ideally, all the nodes in $N$ choose the same chain and attempt to build on top of it. In this case, the stake power of the honest nodes is undivided. However, even if the adversary attempts to divide $N$, it is easy to see that the protocol will not be compromised. The adversary can try to split $N$ by:
    \begin{itemize}
        \item Showing different chains to different subsets of $N$. Say, there is a chain $C$, that was built by $N$ and sent to all in $N$. The adversary will have to show a node, say $n_1$, in $N$, a chain better than $C$, say $C_1$. Now either the node $n_1$ finds the block that leads to its chain being the best chain, or some other node does. If $n_1$ finds it extending $C_1$, it will broadcast the chain since it is an honest node. If any node other than $n_1$ finds it, the protocol executes as usual. So, even if the adversary tries to show a different chain to some node in $N$, $N$ will not be worse off.
        
        \item Giving different block data with the same block header to different subsets of $N$. This case is a trivial subset of the above case. In fact, this can also happen in Ouroboros. If the adversary does try to give two blocks, with the same block header but with different block data, the node which finds the next block that leads to its chain being the best chain will extend the version it has, and all will then accept this.
    \end{itemize}

    \item \emph{Borrow power attack}: Consider the following case, starting from a common block $B$. $N$ sees the block $B_H^1$ as the best block extending $B$. A node $n$ in $N$, is shown a better block $B_A^1$ by the adversary. Now, say, $n$ makes block $B_{n}^2$ extending $B_A^1$, rest of $N$ makes block $B_H^2$ extending $B_H^1$, and the adversary makes block $B_A^2$ extending $B_A^1$; and if, $P(B_{n}^2) > P(B_A^2)$ and $P(B_A^1) + P(B_{n}^2) < P(B_H^1) + P(B_H^2)$; then $N$ will use $\{B_H^1,B_H^2\}$, but now the adversary also has $\{B_A^1,B_{n}^2\}$ which is better than it's own $\{B_A^1,B_A^2\}$. So in a sense the adversary has \emph{borrowed} $n$'s power.
    We use simulations to show that this attack does not significantly affect the adversary's ability to violate common prefix. For details, please refer to Section \ref{ap:bpa}.
    
\end{itemize}


\paragraph{Missing block data attack} Since the nodes are required to build on the chain with the maximum power. The adversary (or any node not in $N$), when it has the best chain, could show its block header to nodes in $N$ and then not send the block data. This would result in the nodes in $N$ being unable to proceed with the protocol execution and hence would stall the protocol. All honest nodes, however, will always broadcast the best block header (that they are aware of) along with its block data and will always include the block data of the previous block when extending it.

To avoid this scenario, we allow nodes to extend blocks without their block data. We call such a block that has only the block header and no block data, as a \emph{null block}. This approach would keep the protocol from stalling as well as prevent the honest chain from losing power (as we can use null blocks when calculating chain power). The information that the previous block was a null block or not is given in the header of the block extending it. Since this information is immutable, given a chain, each block can be unambiguously determined to be null or not. Note that null blocks are different from blocks that have $0$ transactions in their block data.
    
We believe the effect of this attack can be easily mitigated through reward schemes. E.g., by reducing the utility (block reward and transaction fees) of both the owner (publisher) of the null block as well as of the node extending it. As the cost of publishing null blocks is discouraging, the adversary will not launch this attack. The optimal method that can be used to deal with this scenario is dependent on the specifics of the implementation. We, however, assume that this affects neither $tps$ nor chain growth.

\subsection{Borrow Power Attack}
\label{ap:bpa}
Let us calculate the impact of the borrow power attack. Each time the adversarial chain is better than or equal to the honest chain, the adversary can show this chain to a fraction of the honest nodes in $N$ and have them build on it in that slot. When the adversary shows its chain, it may gain or lose utility, depending on the stochastic outcome. Before the adversary shows its chain, it knows the value of power by which it beats the honest chain, as well as the power of its block in that slot. Using this information, the adversary calculates the optimum fraction of nodes to show its chain to. Here, we show that even though the expected utility is positive for the adversary, the gain is, in fact, negligible and marginally affects the adversary's ability to violate common prefix.

\begin{figure}
\centering
\includegraphics[scale=0.30]{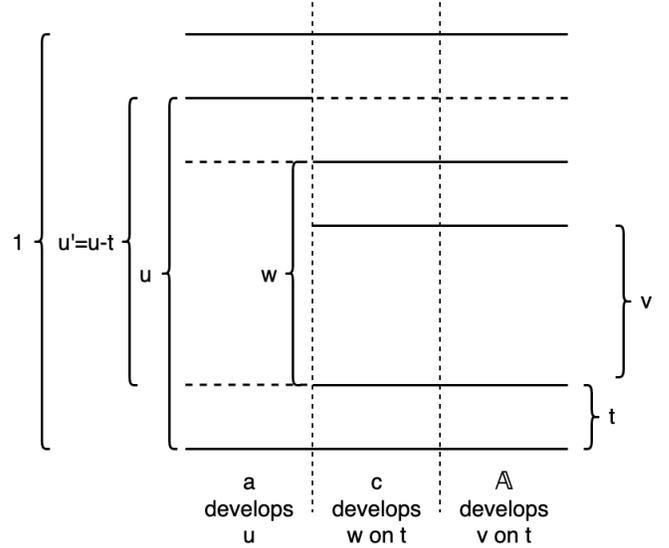}
\caption{Graphical representation of power of blocks in the borrow power attack} \label{fig:bp_tvuw}
\end{figure}

Let us consider the power of the chain of the honest nodes, up till the given slot, to be the baseline. Let us say that the adversarial chain has $t$ more power than that of the honest nodes. Let $v$ be the power of the adversarial block in this slot. We spilt stake power of $\mathbb{H}$ as $c+a$ where $c$ is the fraction of stake power of the nodes in $\mathbb{H}$ that build on the chain that the adversary shows to them (adversarial chain) and $a$ is the fraction of stake power of the remaining nodes in $\mathbb{H}$ which build on the chain built by the honest nodes. Here, we overload the notation $c,a$ to also represent the corresponding sets of nodes, respectively. Let $u,w$ denote the block powers of the blocks generated by $a$ and $c$ respectively, and hence $u$ and $w$ are random variables with \emph{pdfs} $au^{a-1}$ and $cw^{c-1}$ respectively.
Here, we ignore the case $v+t>=1$, as in that case, the adversary can never expect to gain anything and hence will never play that scenario out.
Please refer to Fig. \ref{fig:bp_tvuw} for graphical representation.
Note that all these variables have the constraints:
$0<v<1$, $0<t<1$, $0<u<1$, $0<w<1$, $a>0$, $c>0$, $v+t<1$, $a+c=\alpha_h$.

\begin{figure*}[!t]
\begin{subfigure}{0.32\textwidth}
\centering
\includegraphics[scale=0.4]{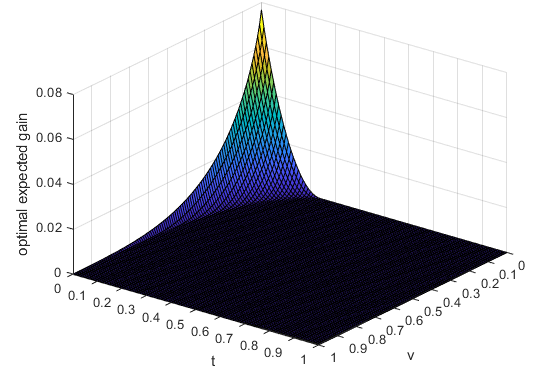}
\caption{Optimal expected gain given $v$, $t$, $r_a=0.45$} \label{fig:g_vt}
\end{subfigure}
\begin{subfigure}{0.32\textwidth}
\centering
\includegraphics[scale=0.4]{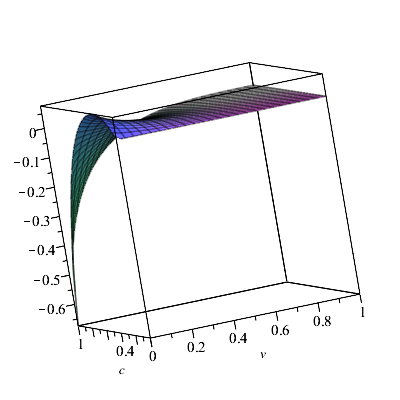}
\caption{Expected gain given $c$, $v$, $r_a=0.45$, $t=0$} \label{fig:g_cv_t0}
\end{subfigure}
\begin{subfigure}{0.32\textwidth}
\centering
\includegraphics[scale=.4]{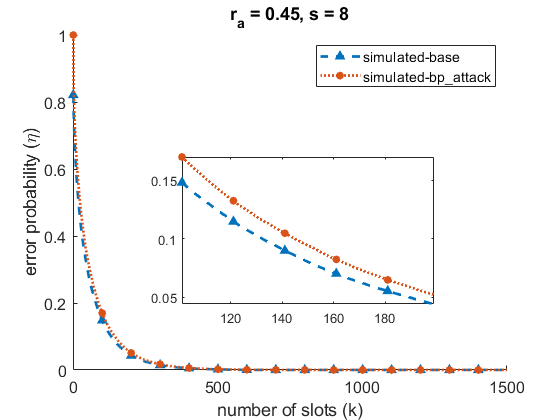}
\caption{Effectiveness of the borrow power attack} \label{fig:bp_error}
\end{subfigure}
\caption{Borrow Power Attack -- Simulation Analysis}
\end{figure*}

Now, as mentioned above, the exact utility of the adversary depends on the power generated by $a$ and $c$. The adversary selects the $c$ that gives the optimal expected utility. The utility here refers to the gain in chain power that the adversary obtains when it conducts the attacks in the given slot.
The change in the difference in chain power of $N$ and the adversary is realized by adopting (or being forced to adopt a different chain than the one held before the attack). Whenever $c$ develops a chain that is better than $a$'s, $N$ adopts it. Since the adversary needs to maintain a chain that differs from the honest chain by at least $k$ blocks, from the one $N$ holds, the adversary can no longer build on the chain that $c$ has developed or the one that it showed to $c$ (unless the adversary wants to start the fork from another point). However, the adversary can now use the chain $N$ has discarded, and it will do so if profitable.

We use the following assumptions:\\
    - We assume that the adversary is comprised of several small nodes. This assumption implies that the adversary has several chains close to the power of its best chain, and when $N$ adopts its chain, it does lose any chain power. This means that whenever $c$ makes a chain better than the one of $a$, and the honest nodes adopt $c$'s chain (the one built on the adversary's best chain before the attack), thus rendering $c$'s chain useless for the adversary (for the current forking attempt), the adversary itself does not lose any chain power.\\
    - For simplicity, we assume that all the stake power in $N$ is held by just one node. That is, there is only one block published by $N$ in any slot. However, in the first slot of the forking attempt, we ignore this assumption. Instead, we assume that $N$ is comprised of several small nodes in favor of the adversary. The adversary can choose between any of the blocks it has built and any block the honest nodes have published except the best block of $N$. Due to this, the adversary, in the first slot of the forking attempt, always has a better or an equivalent block as compared to $N$.

We consider the following six cases. For each of these cases, we now determine the gain in chain power, i.e., the utility gained by the adversary on $N$ when it launches a borrow power attack. Note that this gain is as compared to the case where the adversary does not launch such an attack. The utility gained by $N$ is utility lost by the adversary.
\begin{itemize}
    \item Case 1: $v+t>w+t>u$
    
        If $w>u$, then $N$ gains utility $t$.
        If $u\geq w$, then $N$ gains utility  $(w+t-u)$.
        
        $N$'s expected gain:
        
        \scriptsize
        $
        c1_h = \int_{0}^{v}\!c{w}^{c-1} \left( \int_{0}^{w}\!ta{u}^{a-1}\,{\rm d}u+
        \int_{w}^{w+t}\! \left( w+t-u \right) a{u}^{a-1}\,{\rm d}u \right) 
        \,{\rm d}w 
        $
        \normalsize
        
    \item Case 2: $v+t>u>w+t$; No one gets any utility.
        
    \item Case 3: $u>v+t>w+t$; No one gets any utility.
        
    \item Case 4: $w+t>v+t>u$
        
        If $w>u$, then $N$ gains utility $t$.
        If $u\geq w$, then $N$ gains utility $(w+t-u)$.

        $N$'s expected gain:
        
        \scriptsize
        $
        c4_h = \int_{v}^{v+t}\!a{u}^{a-1} \left( \int_{u}^{1}\!tc{w}^{c-1}\,{\rm d}w
         + \int_{v}^{u}\! \left( w+t-u \right) c{w}^{c-1}\,{\rm d}w \right)
        \,{\rm d}u 
        $
        \normalsize
        
    \item Case 5: $w+t>u>v+t$
        
        If $w>u$, then $N$ gains utility $t$.
        If $u\geq w$, then $N$ gains utility $(w+t-u)$.
        
        Adversary gains utility $(u-(v+t))$

        Adversary's expected gain: $c5_a$
        \scriptsize
        \begin{flalign*}
        & = \int_{v}^{1-t}\!c{w}^{c-1} \left(\int_{v+t}^{w+t}\! \left( u-v-t \right) a{u}
        ^{a-1}\,{\rm d}u \right)\,{\rm d}w \\
         & \qquad + \int_{1-t}^{1}\!c{w}^{c-1} \left(\int_{v+t}^{1}\!
         \left( u-v-t \right) a{u}^{a-1}\,{\rm d}u \right) \,{\rm d}w &
        \end{flalign*}
        \normalsize

        $N$'s expected gain:
        
        \scriptsize
        $
        c5_h = \int_{v+t}^{1}\!a{u}^{a-1} \left( \int_{u-t}^{u}\! \left( w+t-u
         \right) c{w}^{c-1}\,{\rm d}w+\int_{u}^{1}\!tc{w}^{c-1}\,{\rm d}w
         \right) \,{\rm d}u 
        $
        \normalsize

    \item Case 6: $u>w+t>v+t$
        
        Adversary gains utility $w-v$.

        Adversary's expected gain:
        
        \scriptsize
        $
        c6_a = \int_{v}^{1-t}\!c{w}^{c-1} \left(\int_{w+t}^{1}\! \left( w-v \right) a{u}^{a-
        1}\,{\rm d}u \right)\,{\rm d}w
        $
        \normalsize
        
\end{itemize}

So the expected gain of the adversary is,

$f_{eag}(a,c,v,t)=(c5_a+c6_a) - (c1_h+c4_h+c5_h)$

Now, to calculate the optimal $c$, the adversary will do the following:
\begin{enumerate}
    \item Substitute $a=\alpha_h-c$ in $f_{eag}$.
    \item Now, find the $c(v,t)$ that maximizes $f_{eag}(c,v,t)$.
    \item Substitute $v$ and $t$, in $c(v,t)$, to get the optimal value of $c$.
\end{enumerate}

Refer to Fig. \ref{fig:g_vt}, to see how the maximum (over $c$) expected value of $f_{eag}$ changes with $v$ and $t$, and \ref{fig:g_cv_t0}, to see how the value of $f_{eag}$ changes with $c$ and $v$, when $t=0$.

As we see in Fig. \ref{fig:bp_error}, the borrow power attack does not significantly affect the adversary's power to violate finality.

\section{A Comparative Note on v1, Praos and Algorand}
\label{ssec:cnv1pa}
In this section, we compare \proto\ with v1, Praos and Algorand.

The first version of the Ouroboros protocol v1 requires synchrony using NTP. Additionally, it assumes that the adversary can not corrupt an honest node in the time span of an epoch, which could last for a few hours or a few days. Thus, it is not secure against FACs, i.e., against a dynamic adversary. This is due to the fact that all the block publishers of an epoch are common knowledge at the start of the epoch.

The next version of the Ouroboros protocol, Praos, provides security against FACs. Praos does not improve upon the performance of Ouroboros v1. Praos achieves liveness and persistence under what the authors refer to as \emph{semi-synchronicity} (using NTP). However, in Praos, for security, one must set the parameter $f$ appropriately, which implicitly requires prior knowledge of the block propagation delay $\tau$ (as well as $t_{sl}$).
The later versions of Ouroboros, \emph{Genesis} and \emph{Crypsinous}, further add features to the Ouroboros protocol but do not improve upon the performance of Ouroboros v1.

Another popular PoS-based protocol, Algorand \cite{gilad2017algorand}, is independent of the Ouroboros family of protocols. It too requires strong synchrony (using NTP) to achieve \emph{liveness} and \emph{final consensus}. However, it requires $\frac{2}{3}$ majority of honest nodes as opposed to $\frac{1}{2}$ as required by \proto, Ouroboros, \btc\, and many others. Due to this strict assumption, we do not consider Algorand to be in the same league and hence, do not present a comparison. Also, the claims of $tps$ achieved in the Algorand paper are based on tightly parameterized simulations, whereas \proto\ has been bench-marked using generous margins on real \btc\ network data. We believe that, when using the parameters used by Algorand, \proto\ too can achieve similar $tps$.

Table \ref{tab:com_f} gives the value of $t_f=k \cdot t_{sl}$ for different values of $r_a$ and $1-\eta$. We have used $\tau=40$ sec for the comparison. When the adversary has a relative stake of 10\%, it can be easily seen that to have the guarantee of 99.9\% on a published block, \proto\ needs to wait only for 4 minutes, whereas v1 needs 10 minutes and Bitcoin needs 50 minutes. It can be seen from Table \ref{tab:com_f}, \proto\ is about 3 times better than v1 and roughly an order of magnitude better than Bitcoin for the same level of guarantees on finality.
In Table \ref{tab:com_tps}, we compare the throughput, and it is clear that \proto\ performs the best in this regard, better than Bitcoin by an order of magnitude.

\end{appendix}

\end{document}